\documentclass[sigconf, authorversion]{acmart}

\usepackage[utf8]{inputenc} 
\usepackage[T1]{fontenc} 

\usepackage[autostyle=true]{csquotes} 

\usepackage[export]{adjustbox}

\usepackage{amssymb}		
\usepackage[normalem]{ulem}	
\usepackage[flushleft]{threeparttable} 
\usepackage{booktabs,caption}	

\usepackage{subcaption}	
\usepackage{graphicx}

\usepackage{tabularx} 
\usepackage{longtable} 
\usepackage{rotating} 

\usepackage{listings} 

\usepackage{todonotes}

\usepackage{enumitem}

\usepackage{tcolorbox}

\usepackage{xspace}

\usepackage{ulem}

\usepackage{comment}

\RequirePackage[normalem]{ulem} 
\RequirePackage{color}\definecolor{RED}{rgb}{1,0,0}\definecolor{BLUE}{rgb}{0,0,1} 

\newenvironment{myquote}[1]%
  {\list{}{\leftmargin=#1\rightmargin=#1}\item[]}%
  {\endlist}

\AtBeginDocument{%
  }

\copyrightyear{2025}
\acmYear{2025}
\setcopyright{rightsretained}
\acmConference[Australasian Computing Education Conference]{Submitted to ACE 2025}{February 2025}{Brisbane, Australia}
\acmBooktitle{ACE '25}
\acmDOI{}
\acmISBN{xxx-x-xxxx-xxxx-xxxxxxx}

\begin{document}




\title[Oversight in Action]{Oversight in Action: Experiences with Instructor-Moderated LLM Responses in an Online Discussion Forum}



\author{Shuying Qiao}
\email{sqia181@aucklanduni.ac.nz}
\orcid{123}
\affiliation{%
  \institution{University of Auckland}
  \city{Auckland}
  \country{New Zealand}
}

\author{Paul Denny}
\email{paul@cs.auckland.ac.nz}
\orcid{0000-0002-5150-9806}
\affiliation{%
  \institution{University of Auckland}
  \city{Auckland}
  \country{New Zealand}
}

\author{Nasser Giacaman}
\email{n.giacaman@auckland.ac.nz}
\orcid{0000-0001-6885-1571}
\affiliation{%
  \institution{University of Auckland}
  \city{Auckland}
  \country{New Zealand}
}


\newcommand{\llm}{Large Language Model\xspace}
\newcommand{\llms}{Large Language Models\xspace}
\newcommand{\Genai}{Generative AI\xspace}
\newcommand{\genai}{generative AI\xspace}
\newcommand{\ed}{Ed Discussion\xspace}
\newcommand{\hgpt}{HASHTAG-GPT4\xspace}
\newcommand{\mybrand}{\textsc{MyCoolIdeaApproachBrandName}\xspace}
\newcommand{\OOP}{Object-Oriented Programming\xspace}
\newcommand{\Oop}{Object-oriented programming\xspace}
\newcommand{\oop}{object-oriented programming\xspace}
\newcommand{\aida}{AIDA\xspace}

\begin{abstract}

The integration of large language models (LLMs) into computing education  offers many potential benefits to student learning, and several novel pedagogical approaches have been reported in the literature.  However LLMs also present challenges, one of the most commonly cited being that of student over-reliance. 
This challenge is compounded by the fact that LLMs are always available to provide instant help and solutions to students, which can undermine their ability to independently solve problems and diagnose and resolve errors.
Providing instructor oversight of LLM-generated content can mitigate this problem, however it is often not practical in real-time learning contexts.  
Online class discussion forums, which are widely used in computing education, 
present an opportunity for exploring instructor oversight because they operate asynchronously.
Unlike real-time interactions, the discussion forum format aligns with the expectation that responses may take time, making oversight not only feasible but also pedagogically appropriate.
In this practitioner paper, we present the design, deployment, and evaluation of a `bot' module that is controlled by the instructor, and integrated into an online discussion forum. The bot assists the instructor by generating draft responses to student questions, which are reviewed, modified, and approved before release. Key features include the ability to leverage course materials, access archived discussions, and publish responses anonymously to encourage open participation.  We report our experiences using this tool in a 12-week second-year software engineering course on object-oriented programming.  
Instructor feedback confirmed the tool successfully alleviated workload but highlighted a need for improvement in handling complex, context-dependent queries.
We report the features that were viewed as most beneficial, and suggest avenues for future exploration.



\end{abstract}



\maketitle

\providecommand{\DIFadd}[1]{{\protect\color{blue}\uwave{#1}}} 
\providecommand{\DIFdel}[1]{{\protect\color{red}\sout{#1}}}                      
\providecommand{\DIFaddbegin}{} 
\providecommand{\DIFaddend}{} 
\providecommand{\DIFdelbegin}{} 
\providecommand{\DIFdelend}{} 
\providecommand{\DIFaddFL}[1]{\DIFadd{#1}} 
\providecommand{\DIFdelFL}[1]{\DIFdel{#1}} 
\providecommand{\DIFaddbeginFL}{} 
\providecommand{\DIFaddendFL}{} 
\providecommand{\DIFdelbeginFL}{} 
\providecommand{\DIFdelendFL}{} 

\section{Introduction}

The rapid rise and adoption of generative AI, particularly large language models (LLMs), has  started to transform computing education~\cite{denny2024computing}.  A large body of literature exploring the capabilities of LLMs has emerged, highlighting impressive performance for solving programming problems~\cite{finnieansley2022robots, savelka2023gpt3, cipriano2023gpt3}, explaining code and errors~\cite{macneil2023experiences, balse2023evaluating}, and even generating high-quality learning resources~\cite{sarsa2022automatic, jury2024evaluating}. However, the integration of LLMs into educational contexts comes with challenges. A common concern among instructors is that students may become overly reliant on LLMs, potentially diminishing their critical thinking skills and independent learning~\cite{prather2023robots}.  For instance, Prather et al. explored how novice programmers interact with both GitHub Copilot and ChatGPT during problem solving~\cite{prather2024widening}.  They found that while some students effectively leveraged the tools, others who struggled experienced persistent and new metacognitive challenges, including an illusion of competence, highlighting the need for better scaffolding and support.  Concerns around over-reliance are compounded by the fact that LLMs are always available and so it can be tempting for students to immediately seek a solution before thinking more deeply about the problem or exploring other approaches to resolve their issues.  In addition, LLMs are not designed to provide educationally relevant responses and so may reveal direct solutions to students even when that is not desired by the course instructor.

In response, various approaches have been explored to leverage these models in ways that align more closely with educational objectives. One particular strategy has been the development of digital teaching assistants, designed to provide real-time support without revealing complete code solutions to students~\cite{liffiton2024codehelp, kazemitabaar2024codeaid, molina2024leveragingllmtutoringsystems}.
Early evidence suggests that this kind of guardrailed approach is valued by instructors and students alike.  For example, in their exploration of the CodeAid tool, Kazemitabaar et al. found that instructors valued the way the tool prompted students to think critically rather than just offering direct answers~\cite{kazemitabaar2024codeaid}.  In similar work involving a digital teaching assistant for programming, Denny et al. found that students expressed a preference for support that guided them through problem solving steps rather than revealed direct solutions~\cite{denny2024desirable}.  Nevertheless, even with effective guardrails, the immediacy of receiving feedback via digital teaching assistants may encourage poor help seeking practices.  In addition, it is not possible to define guardrails that are perfect in all situations and that consistently generate responses desirable to an instructor.

Including the instructor `in the loop', and allowing them to drive and oversee the process, may help to overcome these challenges~\cite{liu2024hita}.  A good example of this instructor-in-the-loop approach was recently described by Wang et al.~\cite{wang2024tutorcopilothumanaiapproach}.  Their Tutor CoPilot system supports tutors in real-time by providing expert-like guidance tailored to students' needs.  They report the results of a randomised controlled trial which demonstrated that students working with tutors who had access to Tutor CoPilot significantly improved their learning, especially when working with less-experienced tutors.

Class discussion forums are widely used in computing education as an outlet for students to seek help~\cite{vellukunnel2017deconstructing, pieterse2011student}.  Given that responses are  typically generated by peers or the instructor, it is not usually an expectation that responses on class forums are immediate.  Nevertheless, discussion forums can place a burden on instructors---indeed, more than 20 years ago Sheard et al. warned of the impact of their recently adopted web-based discussion forums on the workloads of the staff required to manage them~\cite{sheard2003webbased}.   Thus, it is interesting to consider the use of LLMs as a way to reduce instructor workload when responding to questions on class discussion forums. 

In this paper, we describe the development and deployment of an instructor-driven LLM-bot integrated into an online course discussion forum. The system was designed to assist instructors by generating draft responses to student questions, which could then be reviewed, edited, and released. Key features of the system include the ability to incorporate specific course materials into responses, access archives from previous years, and publish answers anonymously (i.e., not signalled as being generated by the LLM-bot) to foster participation with students.  We deployed this system in a large, second-year software engineering course on object-oriented programming. Through interviews with the course instructor, we investigate how they feel it assisted them and we also explore the specific features of the tool that they used most often.  
These findings offer initial insight into the potential of LLM-based tools for supporting instructors in fostering meaningful peer interactions in online learning environments.




\section{Related work}

\subsection{Discussion Forums and Chatbots}

Online discussion forums have been enthusiastically adopted across many disciplines, as they offer a convenient and asynchronous space for students and instructors to interact, share knowledge, and foster collaborative learning opportunities~\cite{ackerman2021synchronous, khoshnevisan2021developing}.  In computing education, where online discussion forums are commonly used, there is evidence that meaningful interaction with forums can aid learning. Research by Newby et al. found that students actively posting questions, especially those whose posts were highly readable (with respect to a readability metric) and received peer engagement, performed better in a CS1 course than students who did not post~\cite{newby2024askers}.  Similarly, Campbell et al. investigated success factors in an online introductory programming course and discussed the important role of online discussion forums for creating a sense of community and support among students~\cite{campbell2016factors}. They found that high engagement on the forums, marked by thousands of student and instructor posts, was critical to maintaining a sense of belonging and motivation, contributing to overall student satisfaction and perceived connection to the course.  

Although student engagement is clearly an important factor, it does not always happen automatically and appropriate instructor facilitation such as providing regular feedback is also important for fostering productive discussions~\cite{nandi2012factors}.  This presents a challenge, as managing online forums can be time consuming for instructors, especially with large classes. Previous research highlights the increased workload placed on instructors managing discussion boards~\cite{sheard2003webbased}. While students benefit from frequent instructor feedback, balancing workload while maintaining engagement is an ongoing challenge~\cite{wang2022comparative}.

In recent years, AI-powered chatbots and discussion forums have emerged as potential solutions to alleviate instructor workload by automating responses to routine questions. Such chatbots can offer timely feedback, provide students with hints instead of direct answers, and foster engagement through conversational interaction~\cite{khoshnevisan2021developing, wambsganss2021arguetutor}.  For example, Liu et al.  describe the integration of a suite of AI tools, including an AI-powered chatbot called ``CS50 Duck'', within Harvard University's Introduction to Computer Science course~\cite{liu2024teaching}. This tool was designed with `pedagogical guardrails' to avoid simply providing solutions, instead guiding students toward problem-solving approaches.  In similar work, Zamfirescu-Pereira et al. describe the deployment of a GPT-4-based AI assistant, ``61A Bot'', within a large introductory CS course at UC Berkeley~\cite{zamfirescupereira2024botreport}. This assistant was designed to provide students with immediate, contextualised feedback on coding assignments directly within their command-line autograder interface. The bot was designed to offer students debugging hints and concept reinforcement without providing solutions. Data from over 100,000 interactions demonstrated a notable decrease in student posts on the course forum and faster homework completion times. Although student feedback was generally positive, highlighting the bot's helpfulness, there were occasional issues with vague or overly specific feedback.  In addition, the limitations discussed by the authors included reduced visibility for instructors into student challenges.  While it is feasible to design bots that often produce high-quality responses on discussion forums, especially using techniques such as retrieval-augmented generation (RAG) as described by Hicke et al.~\cite{hicke2023aitaintelligentquestionanswerteaching}, allowing instructor oversight can address challenges around occasional inappropriate feedback and provide instructors with insight into the kinds of issues that students are frequently encountering.

\subsection{LLMs in Computing Education}

Large language models (LLMs) have created new opportunities in computing education, from the generation of learning resources~\cite{Sarsa2022, macneil2023experiences} to powering new kinds of pedagogical approaches~\cite{denny2024prompt, denny2024explaining}.   Much recent work has begun to explore the integration of LLM-powered tools into computing classrooms and has shown great promise.  For example, Jury et al. explored LLM-generated worked examples in an introductory programming course, finding that students perceived the step-by-step guidance as highly beneficial for understanding complex concepts and reinforcing their learning~\cite{jury2024evaluating}.
Liffiton et al. built a LLM-powered tool called CodeHelp to provide assistance to programming students~\cite{liffiton2023codehelp}.  When stuck, students could enter the programming language, code, error message and their issues. The tool used a series of prompting strategies to generate helpful output, which most students perceived as useful.

However, integrating LLMs into computing education contexts is not without challenges. Prather et al. explored how novice programmers used generative AI tools, such as ChatGPT and GitHub Copilot, when solving programming problems~\cite{prather2024widening}.  Through observations, as well as interviews and eye-tracking, they found that students with higher skill levels effectively used AI to accelerate their work, recognising and ignoring unhelpful suggestions. Conversely, students who struggled often faced persistent and even new metacognitive challenges, such as over-relying on AI, experiencing confusion, and developing an ``illusion of competence'' about their problem-solving abilities. One way to avoid over-reliance is to throttle or limit the extent to which students can seek help.

While LLMs can offer scalable support, their outputs may not always align with instructional goals, requiring human oversight to ensure educational relevance and appropriateness. Instructors must ensure that LLM-generated responses complement rather than replace meaningful instructor-student interactions~\cite{liu2024hita}.  For example, Hellas et al. investigated the effectiveness of OpenAI's Codex and GPT-3.5 in responding to help requests from beginner programming students~\cite{hellas2023exploring}. Using data from an introductory programming course, they examined how well these LLMs identified and addressed issues in student code.  Although GPT-3.5 identified at least one correct issue in 90\% of cases, it often missed multiple issues and generated false positives, leading to potential confusion. Additionally, LLMs frequently included model solutions even when explicitly prompted not to, which could hinder learning.  The authors suggest that, while promising, LLMs should not replace human assistance in educational settings, and this suggests that -- where possible -- there may be value in allowing instructors to review output before it is shown to students.


\section{AIDA Design}

\begin{figure*}[ht]
    \centering
    \includegraphics[width=.8\textwidth]{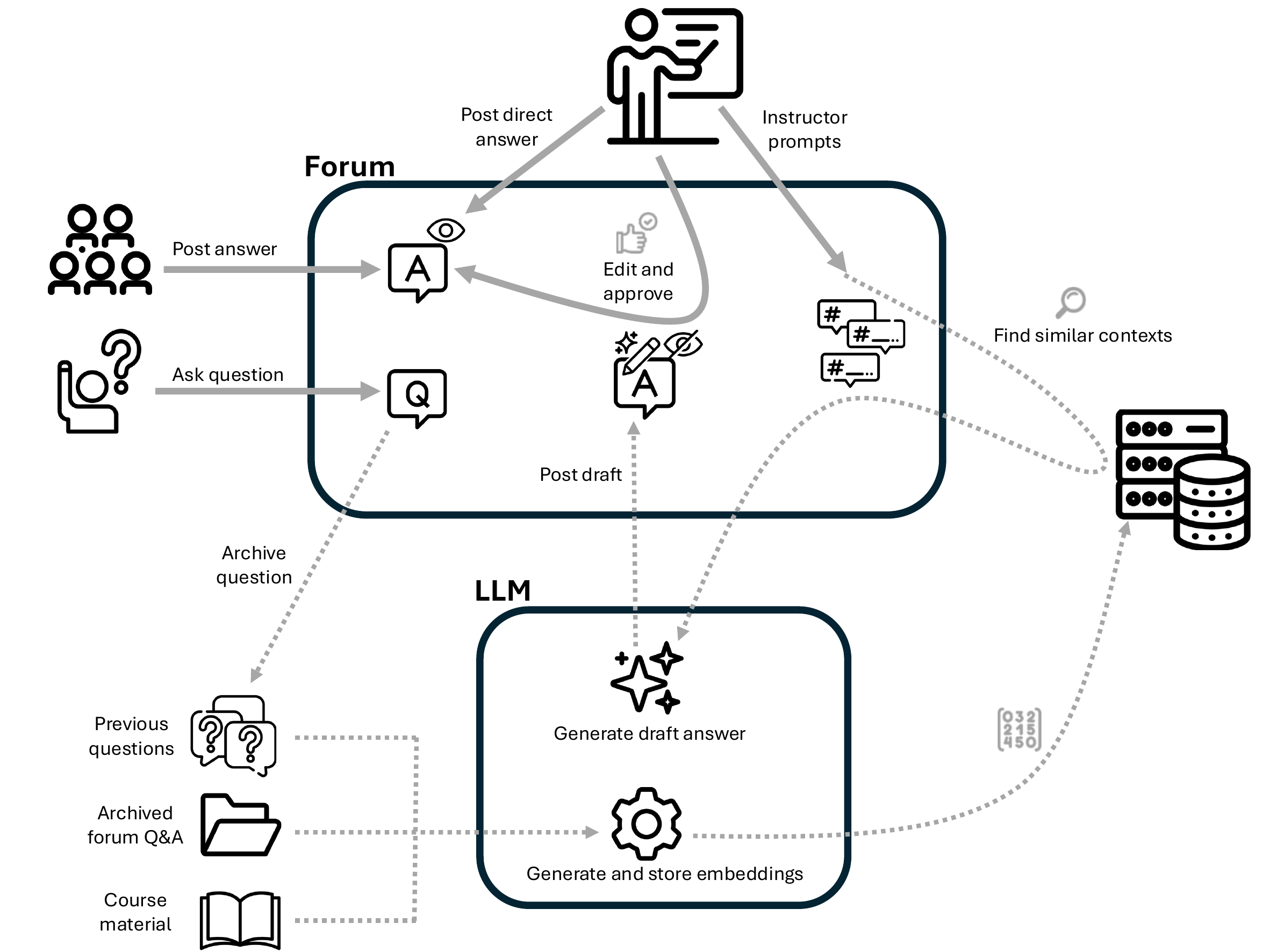}
    \caption{Illustration of \aida's core implementation workflow in assisting instructors. Solid arrows represent direct user actions, such as students posting questions and instructors using \aida prompts to draft answers. Dashed arrows indicate processes handled by \aida, such as managing data embedding, context matching, and generating draft responses.}
    \label{fig:implementation}
\end{figure*}

Integrating AI in educational environments offers opportunities for improving student learning in several ways. 
The \textit{AI Discussion Assistant} (AIDA) system in this project served two roles. The primary role was to draft responses to student queries to support instructors and improve instructional efficiency and effectiveness. 
AIDA could also post anonymously in the discussion forum, a feature that could be leveraged by the instructor for pedagogical reasons.
This was intended to allow students to engage and share ideas freely without knowing the responses were generated by an AI model. This feature encouraged open participation and exploration of ideas.
Approval from the University's Human Participants Ethics Committee was granted (reference \#UAHPEC26871) to ensure transparency about the bot's presence in the course, without disclosing its identity to students. Students were informed about the existence of an AI module integrated in the forum in order to promote informed engagement, while maintaining a genuine learning environment.

\subsection{Theoretical Foundations}


The goal of \aida to assist instructors in drafting responses is grounded in \textit{cognitive load theory (CLT)}~\cite{chandler1991cognitive}.
CLT emphasises the importance of optimising cognitive resources by reducing extraneous load---effort spent on tasks that do not directly contribute to meaningful learning or outcomes. 
By generating draft responses to student queries, \aida 
reduces the extraneous cognitive demands placed on instructors, such as gathering information, organising content, and drafting responses. This allows instructors to allocate more of their cognitive resources to germane load---focusing their expertise on refining and contextualising responses to maximise their pedagogical value. 
Not only does this help reduce the instructor's workload, 
but it also has the potential to improve the quality and timeliness of responses, benefiting students directly.


The secondary role of \aida is to  act anonymously as a student peer in the discussion forum, leveraging principles from \textit{peer learning}~\cite{kear2004peer} and \textit{social constructivism}~\cite{rivera2024learning, sharif2015discussion}. 
By making the AI bot indistinguishable from a fellow student, it is intended to foster a collaborative environment where students feel comfortable exploring and sharing ideas.
This approach addresses a common criticism by students that AI bots lack the intelligence and emotions of human interaction~\cite{chen2023artificial, gallacher2018my}.




\subsection{Technical Workflow}\label{sec:technical_workflow}

\aida is designed to deliver context-aware responses that are closely tied to the course-specific content, using a technique often referred to as retrieval augmented generation. Different from simpler solutions that utilise LLMs ``out of the box'', \aida is provided access to a rich repository of course content, previous questions, and answers from past years. These are preprocessed into vector embeddings to form a structured vector database for quick retrieval during use as shown in Figure~\ref{fig:implementation}. When a student asks a question on the discussion forum, \aida converts this into an embedding and searches for matches in the database to provide relevant context. This allows responses to be drawn from a rich information base that is more accurate and appropriate in the context of the course.

\aida is not meant to be a fully automated bot. Its main goal is to support---not replace---instructors in responding to student queries. For this reason, \aida drafts private responses for instructors to review and refine as they see fit. By having access to relevant context, this workflow improves instructor efficiency. Instructors focus on 
refining
the drafts rather than producing responses from scratch, and having this built-in process eliminates the need for external LLMs to draft responses that would require providing course context manually. Furthermore, instructors might not always be able to recall or access past relevant questions, particularly from previous course offerings, to use as context for drafting responses.

\subsection{Instructor Workflow and Prompts}

\begin{figure}[ht]
    \centering
    \includegraphics[width=0.9\columnwidth]{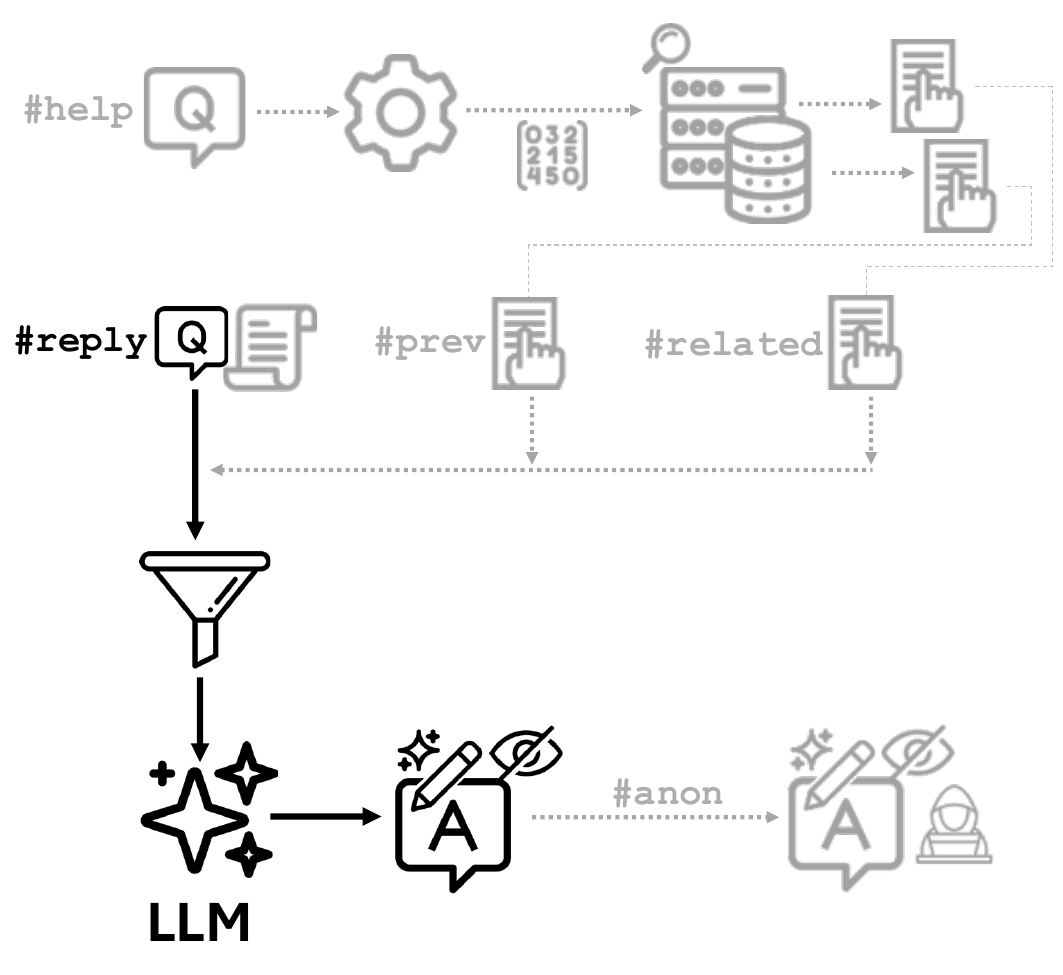}
    \caption{Illustration of the \aida prompts available to the instructor. Instructors can provide additional context and instructions to \aida to generate more tailored responses. Greyed out prompts and components are optional.}
    \label{fig:prompts}
\end{figure}

Figure~\ref{fig:implementation} illustrates \aida's overall high-level architecture, in terms of its implementation workflow and core components. We now provide a closer look at the workflow an instructor would follow when using \aida to draft responses. A key aspect of \aida is that it is designed to work \textit{collaboratively with instructors}, rather than as a standalone tool; although LLMs excel at incorporating explicit text-based context supplied in prompts, they do not have the ``unspoken'' context that uniquely resides with the instructor. Instructors bring an invaluable layer of ``hidden context'' to educational interactions---insights taken from verbal classroom discussions, individual student needs, or other specific pedagogical goals that may not have been explicitly outlined in any static course material. Such hidden context often informs how an instructor might want to tailor their responses to better resonate with students. 


Interaction with \aida  is defined by a set of prompts that the instructor types as comments in the discussion forum. When the comment with an \aida prompt is submitted, \aida recognises these prompts, hides them from students, and performs the necessary action based on the prompt's type. These prompts were crafted to offer transparency and control, and maximise the generated content's alignment with the course's specific context and instructor expectations. When students post questions on the discussion forum, the instructor decides whether to directly respond (without \aida's assistance) or to use \aida to draft a response. If the instructor chooses the latter, then they have the prompts shown in Figure~\ref{fig:prompts} available.

The \texttt{\#reply} prompt is the only required prompt, as it is the one that triggers \aida to draft a response. All of the greyed-out components shown in Figure~\ref{fig:prompts} are optional, and the instructor can choose to skip them if they wish. If they do so, the instructor is effectively directly calling the LLM to generate a response based on the student's question alone. In addition to this bare-minimum usage, the instructor can optionally provide additional instructions specific to the question alongside the \texttt{\#reply} prompt. This might involve asking for specific levels of detail, incorporating particular examples, or explaining with specific learning objectives in mind.

If the instructor wishes to incorporate additional context (in the form of course material or previous discussions) into the response, they will need to discover the most likely relevant contexts. The \texttt{\#help} prompt triggers \aida to search the database of embeddings for the most relevant contexts, which are then presented to the instructor for inspection in two numbered lists (\textit{previous} and \textit{related} context items) with each list consisting of the most relevant five items in that category. \aida does this by converting the student's question into an embedding and searching the database for the highest-matching relevant contexts.

Should the instructor wish to incorporate any of those contexts, they specify the relevant context item(s) (using their index in the list) with the \texttt{\#prev}  and/or \texttt{\#related} prompts (in conjunction with the required \texttt{\#reply} prompt). The \texttt{\#prev} prompt is used to specify specific context(s) referring to previous forum posts (including previous years' archived posts), and will  include the post's question and answers. The \texttt{\#related} prompt is used in a similar manner, but to specify specific course material or assignment handouts.

\aida then generates a draft response based on the provided context and instructions, which the instructor can review and refine before making it visible on the discussion forum. At the time of this study, \aida used OpenAI's GPT-4 model to generate these responses.
Finally, if the instructor would like to post the response anonymously, they can use the \texttt{\#anon} prompt to do so; from the student's perspective, the response would appear as if another student had posted it anonymously.

\section{Evaluation}

\subsection{Participants}

To evaluate AIDA, both from the student and instructor perspective, we deployed it throughout the first six weeks of a 12-week course (SOFTENG 281) on object-oriented programming (OOP) that is compulsory for computer systems engineering, electrical and electronic engineering, and software engineering students at the University of Auckland. The programming language used in this OOP course is Java. The course used \textit{Ed Discussion}\footnote{\url{https://edstem.org}} as the online discussion forum platform, and \aida was deployed on this platform. As Ed Discussion was also used in the previous year for the same course, the discussions from that year were anonymised and made available to \aida.

The participants were the instructor and students in the course.  The \aida system was developed with the objective of both helping instructors respond to student queries, and in turn help students receive high-quality feedback when seeking help.  When answering students' questions, the system would return answers as the instructor (default) or anonymously as a peer (when \texttt{\#anon} was used by the instructor).  Approval for this evaluation was granted by the University of Auckland Ethics Committee (Reference \#UAHPEC26871).

\subsection{Data collected}


\subsubsection{Forum posts}
The data comprised all posts made in the \ed forum in the first six weeks of the course. Our ethics application advised participants that they could withdraw from the study by deleting their posts so that their responses would not be included in the research. There were a total of 277 posts authored by 105 unique students (of the 307 enrolled in the course), and 24 posts were deleted. 
Of the 253 non-deleted posts, the instructor used AIDA to respond to 95 of them.  These 95 questions were used in our research evaluation.
For these 95 posts, we collected data on the frequency of each hashtag prompt used by the instructor, the extent to which the instructor edited the AIDA-generated draft answers (including both content removals and additions), how the instructor utilised \aida when generating responses (specifically which \aida hashtags were used), and whether the instructor adopted and revealed the AIDA-generated answers to students.



\subsubsection{Instructor interview}

To understand the instructor's experience with the \aida system, we conducted an end-of-course interview focused on usability, response quality, and overall satisfaction. Prior to the interview, the researcher analysed the usage of \aida to identify interesting usage decisions. These insights were then used to prompt the instructor during the interview, asking why they made certain decisions and how they felt about the system's performance. The instructor shared insights on the system's ease of use, the relevance and accuracy of responses, and suggestions for improvement.

\section{Results and Discussion}

\subsection{Prompt Usage}

\begin{table}[ht]
    \centering
    \caption{All combinations of hashtags, and their frequencies, used by the instructor in their interactions with AIDA. We separate the usage of \texttt{\#reply} with additional instructions ($\blacksquare$) or without ($\varnothing$). Combinations incorporating \texttt{\#anon} are grey.}

\begin{tabular}{lc}
\hline
\textbf{Prompts Combination}                                         & \textbf{Usage Proportion}   \\ \hline
\#help                                                               & 34\%                        \\
{\color[HTML]{9B9B9B} \#reply$\varnothing$ \#anon}                  & {\color[HTML]{9B9B9B} 21\%} \\
{\color[HTML]{9B9B9B} \#reply$\blacksquare$  \#anon}                & {\color[HTML]{9B9B9B} 18\%} \\
{\color[HTML]{9B9B9B} \#reply$\varnothing$ \#anon \#related}        & {\color[HTML]{9B9B9B} 8\%}  \\
\#reply$\blacksquare$                                               & 6\%                         \\
\#reply$\varnothing$                                                & 6\%                         \\
\#reply$\varnothing$ \#prev                                         & 2\%                         \\
{\color[HTML]{9B9B9B} \#reply$\varnothing$ \#prev \#anon}           & {\color[HTML]{9B9B9B} 2\%}  \\
{\color[HTML]{9B9B9B} \#reply$\varnothing$ \#related \#prev} & {\color[HTML]{9B9B9B} 2\%}  \\
{\color[HTML]{9B9B9B} \#reply$\blacksquare$  \#anon \#related}      & {\color[HTML]{9B9B9B} 1\%}  \\
\#reply$\blacksquare$ \#related \#prev                       & < 1\%                         \\ \hline
\end{tabular}

    \label{table:combined_prompt}
\end{table}

Table~\ref{table:combined_prompt} presents the frequency of hashtag prompt combinations used by the instructor. As indicated in Figure~\ref{fig:prompts}, the \texttt{\#help} prompt is employed to search for contexts in prior posts or course material which could then be used for the \texttt{\#related} and \texttt{\#prev} prompts to draft responses. The \texttt{\#help} prompt was used extensively, accounting for 34\% of the total interactions. This indicates that the instructor frequently sought additional context they believed could be valuable. However, the \texttt{\#related} and \texttt{\#prev} prompts were used about half as often (only 16\% of the time). This suggests that, despite seeking context with \texttt{\#help}, the instructor did not consistently incorporate context from previous posts or material.

In drafting responses, the instructor utilised the \texttt{\#reply} prompt mostly on its own without feeling the need to provide additional instructions or context (\texttt{[\#reply$\varnothing$ \#anon]} and \texttt{[\#reply$\varnothing$]}). At first glance, this suggests that the instructor was comfortable with \aida generating responses based on the student's query alone, without needing to provide additional context or instructions. However, we will see in the instructor interview that the instructor was not always satisfied with the responses generated by \aida when no additional instructions were specified. In fact, the instructor sometimes used the \texttt{[\#reply$\varnothing$ \#anon]} combination as a hack to define an anonymous comment on the discussion forum that they would then go on to edit quite substantially before publishing.

The instructor still extensively used the \texttt{\#reply} prompt while specifying additional instructions, but without using past contexts (\texttt{[\#reply$\blacksquare$ \#anon]} and \texttt{[\#reply$\blacksquare$]}). This combination collectively accounted for 24\% of the prompt interactions. This suggests that the instructor often had a specific idea of what they wanted the response to contain, and did not feel the need to provide additional context when drafting the response. From a practical perspective, it is likely that the instructor saw this as a quick and efficient way to create responses that aligned with their expectations, allowing them to easily make any needed adjustments later.

When considering the drafting of responses that incorporated course-specific contexts, the \texttt{\#related} prompt tended to be used more often than the \texttt{\#prev} prompt (11\% vs. 6\%). Using both of them together was relatively rare (2\%). This suggests that the instructor found more value in incorporating related materials than in incorporating previous posts when drafting responses. Alternatively, it could also suggest that the nature of the questions tended to be more specific to the course materials (which included assessment handouts), than generic questions that were likely to have been asked before in previous iterations of the course.

The diverse usage of prompt combinations, despite some being more frequently used than others, highlights an important insight: the effectiveness of LLM integrations in a discussion forum does not hinge on a single ``best'' method. Instead, it appears valuable to grant instructors with flexibility that empowers them to tailor their engagement based on the nature of each question. As evidenced in the subsequent instructor interview, it becomes clear that response drafting often involved an iterative approach, sometimes experimenting with different strategies to uncover the best generated response for each unique question.

\subsection{Edits to Generated Answers}

\begin{figure}[t]
    \centering
    \includegraphics[width=1\columnwidth]{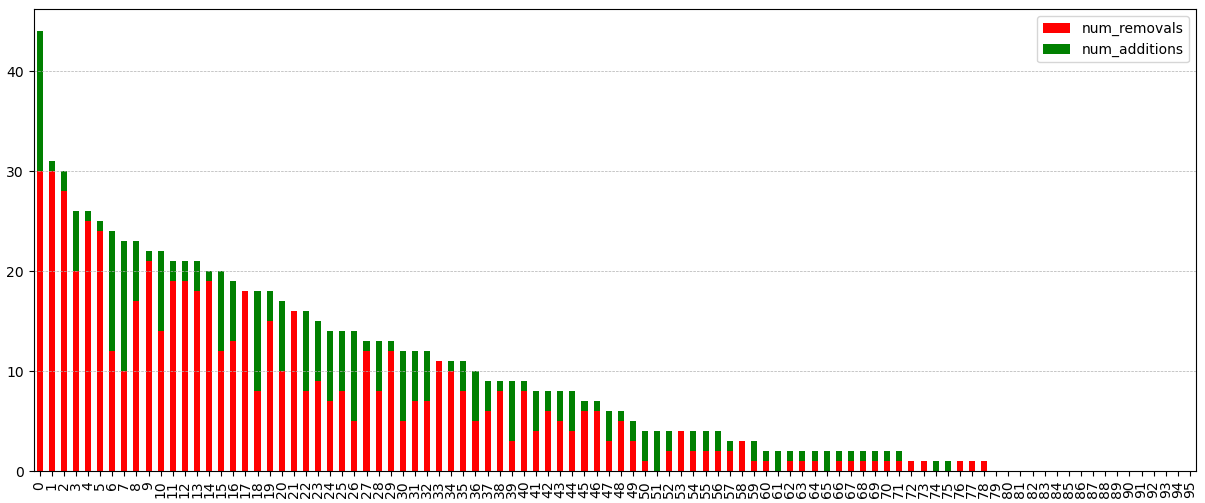}
    \captionof{figure}[Edit information for generated answers]{Number of edits made to all adopted generated answers (removals shown in red; additions shown in green).}
    \label{fig:edit_add_remove}
\end{figure}


When \aida drafts a response, the instructor has the option to edit the response before making it visible on the discussion forum. We used a text differencing tool\footnote{\url{https://www.diffchecker.com}} to calculate the number of edits made by the instructor for each adopted generated answer. The stacked bar chart in Figure~\ref{fig:edit_add_remove} shows the number of edits for each adopted generated answer. Red bars indicate the number of removals from the original answer and green bars indicate the number of additions to the original answer. There are also some answers that were not edited, which means they were adopted by the instructor without any modifications.

Most of the edited answers involved removing content from \aida's drafted answer, with additions being less common. The high proportion of removals for most posts may indicate that the answers the LLM generated contained redundant content, content the instructor felt was inappropriate, or overly detailed. With fewer additions needed overall, it suggests that the instructor felt that the draft answers generally contained the necessary information and only required minor modifications. Of the 95 adopted generated answers, over half required fewer than 10 edits (additions and removals combined), with very few requiring a large number of edits. The instructor interview sheds further light on this.  


\subsection{Example Questions}

\begin{figure}[!htbp]

\begin{tcolorbox}[boxsep=2pt,left=2pt,right=2pt,top=2pt,bottom=2pt]
    \footnotesize
    $
    \begin{array}{l}
    \includegraphics[width=4mm]{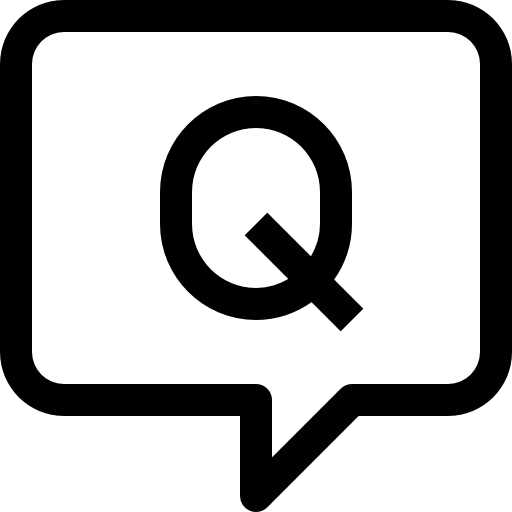}
    \end{array}
    $ \textbf{Student question:}

    \vspace{2mm}

    Hi, I'm having issues at the "Authentication with GitHub" step of the course Git setup instructions.

    \vspace{2mm}

    I'm a bit lost and don't really know how to proceed.

    \vspace{2mm}

    When I enter git commit, I get the following:

    \begin{myquote}{2mm}


        On branch master

        Initial commit

        Untracked files:

         (use ``git add <file>...'' to include in what will be committed)

        Here comes a list of files

        nothing added to commit but untracked files present (use ``git add'' to track)

    \end{myquote}

    \vspace{2mm}

    When I enter git push, I get the following:

    \begin{myquote}{2mm}

        fatal: No configured push destination.

        Either specify the URL from the command-line or configure a remote repository 




    \end{myquote}

    \vspace{2mm}

    \rule[0pt]{230pt}{0.1pt}

    \vspace{2mm}

    $
    \begin{array}{l}
    \includegraphics[width=5mm]{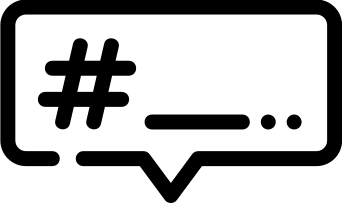}
    \end{array}
    $ \textbf{Instructor prompt combination:}

    \vspace{2mm}

    \textbf{\#anon}
    \textbf{\#reply} Start off by explaining where the possible issue is. Make sure you explain this for a student new to Git and GitHub. The main branch is called "main". Make it clear there's a difference between Git and GitHub. So maybe separate some simple instructions to learn to use Git from a blank folder, and then in a separate section extend this to introduce GitHub.

    \vspace{2mm}

    \rule[0pt]{230pt}{0.1pt}

    \vspace{2mm}

    $
    \begin{array}{l}
    \includegraphics[width=5mm]{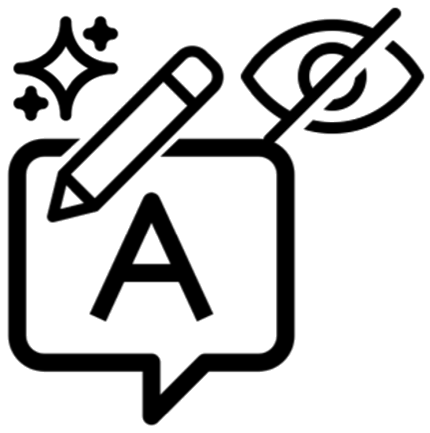}
    \end{array}
    $ \textbf{Generated answer with instructor edits:}

    \vspace{2mm}

    \DIFdelbegin \DIFdel{Understanding the Issue
    }

    \DIFdelend Firstly, before jumping to the issue of authentication with GitHub, you need to understand some basics of Git, a version control system that lets you manage and keep track of your code projects. The master branch that you're on is now more commonly referred to as main.

    \vspace{2mm}

    The error message you received when you entered git commit is Git telling you that there are some files in your current directory that have not been staged for commit. Git only tracks the files that you explicitly tell it to using the git add command. You will need to use git add \DIFdelbegin \DIFdel{<file> }\DIFdelend before git commit.

    \vspace{2mm}

    When you tried git push, it seems there isn't a remote repository (on GitHub) linked to your local repository (on your PC). Hence, the warning about no configured push destination.

    \vspace{2mm}

    \textbf{Getting Started with Git}

    To start practising with Git, open up a blank folder and initialise it as a Git repository using \texttt{git init}.

     \vspace{2mm}

    To stage a file for commit run \texttt{git add <filename>}. You can check what files are staged using git status.

    \vspace{2mm}

    To commit your changes, write \texttt{git commit -m "Your commit message"}.

    \vspace{2mm}

    \textbf{Connecting Git with GitHub}

    GitHub is a remote platform for hosting Git repositories. To push your local repository to GitHub you need to connect the two.

    \vspace{2mm}

    First, create a new repository on GitHub. Then, link it with your local repository using \texttt{git remote add origin URL\_OF\_YOUR\_GITHUB\_REPO}.

    \vspace{2mm}

    You then upload your commits to GitHub with \texttt{git push -u original main}.

    \vspace{2mm}

    \texttt{// Commands \\
    git init \\
    git add <filename> \\
    git commit -m "Commit message" \\
    git remote add origin <URL\_OF\_YOUR\_GITHUB\_REPO> \\
    git push \DIFdelbegin \DIFdel{origin main}}
    \DIFdelend \DIFaddbegin \DIFadd{\texttt{ -u origin main}
    }\DIFaddend

    \vspace{2mm}

    Finally, the authentication issue seems like a separate topic, it's necessary to be authenticated in order for Git to interact with GitHub on your behalf. You can authenticate \DIFdelbegin \DIFdel{using SSH or by }\DIFdelend using the token-based system.

\end{tcolorbox}
    \caption{Due to the generic nature of this question, the instructor did not search for similar contexts using \texttt{\#help}. Instead, they used the \texttt{\#reply} prompt with additional instructions to provide a comprehensive and structured answer.}
    \label{fig:example-1}
\end{figure}

In this section, we illustrate several questions from the course and show how the instructor used the prompts in different ways. Figure~\ref{fig:example-1} shows an error the student encountered while setting up Git.
The instructor decided this query could be easily answered without the need for course-specific context.
After noticing particular misconceptions, the instructor crafted additional instructions for the \texttt{\#reply} prompt to guide the LLM's output. The instructor opted to use the \texttt{\#anon} prompt to give the impression that a helpful anonymous peer had answered, hoping that this would encourage a sense of community among the students to support each other.

\begin{figure}[!htbp]

\begin{tcolorbox}[boxsep=2pt,left=2pt,right=2pt,top=2pt,bottom=2pt]
    \footnotesize
    $
    \begin{array}{l}
    \includegraphics[width=4mm]{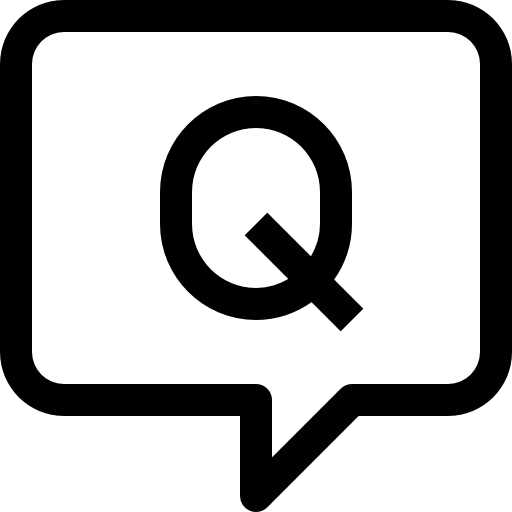}
    \end{array}
    $ \textbf{Student question:}

    \vspace{2mm}

    Hello,

    Could the recording for the lecture please be posted on Panopto, I was sick and could not attend the lecture.

    Thank you

    \vspace{2mm}

    \rule[0pt]{230pt}{0.1pt}

    \vspace{2mm}

    $
    \begin{array}{l}
    \includegraphics[width=5mm]{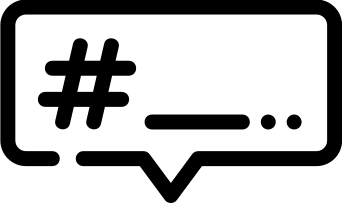}
    \end{array}
    $ \textbf{Instructor prompt combination:}

    \vspace{2mm}

    \textbf{\#help}

    \vspace{2mm}

    \textbf{\#reply \#prev 2 292 473}

    \vspace{2mm}

    \rule[0pt]{230pt}{0.1pt}

    \vspace{2mm}

    $
    \begin{array}{l}
    \includegraphics[width=5mm]{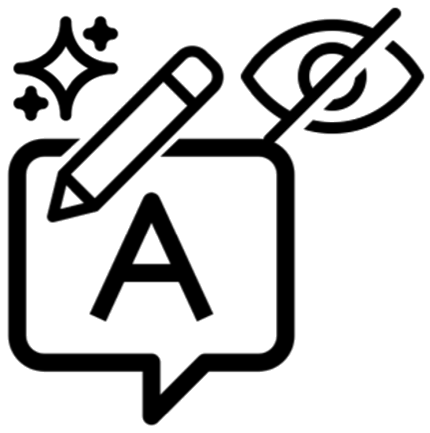}
    \end{array}
    $ \textbf{Generated answer with instructor edits:}

    \vspace{2mm}

    Panopto will have the lecture recording available soon, keep an eye out for it.

\end{tcolorbox}

    \caption{Example of a course-related administrative question. After using the \texttt{\#help} prompt, the instructor selected the three relevant previous contexts to generate a response.}
    \label{fig:example-2}
\end{figure}

The next example in Figure~\ref{fig:example-2} is a question related to the timing of the release of a lecture recording. The instructor suspected that this question has been asked before, so sought to find the relevant context using the \texttt{\#help} prompt. This reveals identifiers for the most-likely related context, which the instructor then used to generate a response using the \texttt{\#prev} hashtag. The generated answer was concise and to the point, providing the information the student needed. The instructor felt that no edits were necessary.

\begin{figure}[!htbp]

\begin{tcolorbox}[boxsep=2pt,left=2pt,right=2pt,top=2pt,bottom=2pt]
    \footnotesize
    $
    \begin{array}{l}
    \includegraphics[width=4mm]{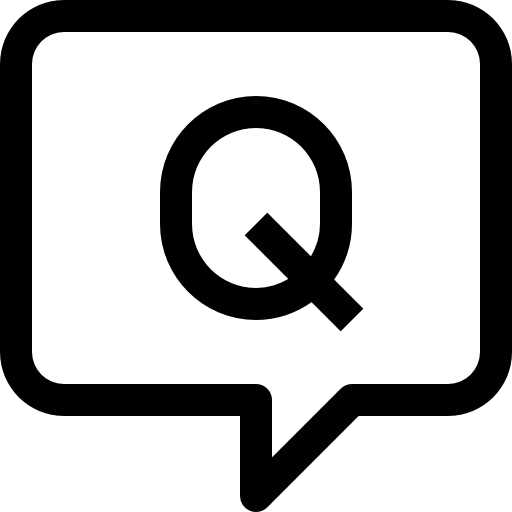}
    \end{array}
    $ \textbf{Student question:}

    \vspace{2mm}

    can we assume that the bookings will be made in order?

    like, 10/10/2024, then 11/10/24, etc

    \vspace{2mm}

    \rule[0pt]{230pt}{0.1pt}

    \vspace{2mm}

    $
    \begin{array}{l}
    \includegraphics[width=5mm]{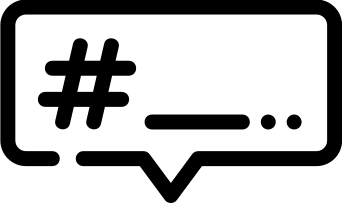}
    \end{array}
    $ \textbf{Instructor prompt combination:}

    \vspace{2mm}

    \textbf{\#help}

    \vspace{2mm}

    \textbf{\#reply \#related 42,44 \#anon}

    \vspace{2mm}

    \rule[0pt]{230pt}{0.1pt}

    \vspace{2mm}

    $
    \begin{array}{l}
    \includegraphics[width=5mm]{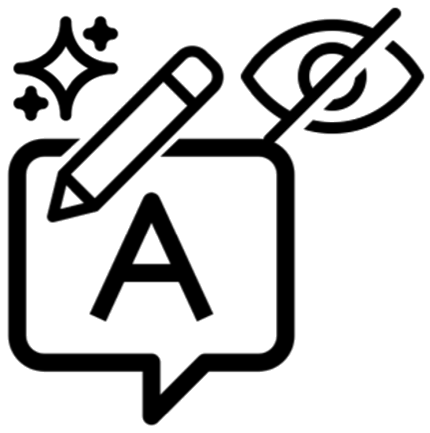}
    \end{array}
    $ \textbf{Generated answer with instructor edits:}

    \vspace{2mm}

    \DIFdelbegin \DIFdel{In object-oriented programming and Java particularly, we usually don}\DIFdelend \DIFaddbegin \DIFadd{No, you shouldn}\DIFaddend 't make assumptions about the order of operations like venue bookings\DIFdelbegin \DIFdel{. Instead, we design programs to handle bookings in whatever order they come in.
        }

        \DIFdel{Here's a simplistic example using ArrayList:
        }

        \DIFdel{import java.util.ArrayList;
        import java.util.Collections;
        }

        \DIFdel{public class Main }
        \\
        \DIFdel{public static void main(String}
        \DIFdel{ args) }
        \\
        \DIFdel{\texttt{ArrayList<String> bookings = new ArrayList<String>();}
                \\
                \texttt{bookings.add("10/10/2024");}
                \\
                \texttt{bookings.add("11/10/2024");}
                \\
                \texttt{bookings.add("9/10/2024");}
        }

        \DIFdel{Collections.sort(bookings); // Sorts the bookings in ascending order.
        }

        \DIFdel{for (String booking : bookings) }
        \\
        \DIFdel{System.out.println(booking);
                }

        \DIFdel{In this code, we're adding bookings out of order, then we sort them}\DIFdelend .

\end{tcolorbox}

    \caption{Example of an assignment-specific question. The instructor used the \texttt{\#help} prompt and selected two related previous contexts to generate a response anonymously. The draft answer was significantly edited by the instructor to remove excess information and provide a more direct response.}
    \label{fig:example-3}
\end{figure}

The final example in Figure~\ref{fig:example-3} is a question related to one of the course assignments. After using the \texttt{\#help} prompt, the instructor found similar contexts and used the \texttt{\#related} prompt to generate a response.
The LLM generated a lengthy draft answer that included a code example. However, given that the question was only seeking clarification for a detail of the assignment requirements, this code was not necessary and was deleted by the instructor.

\subsection{Instructor Feedback}


The interview with the instructor focused on three main aspects: the usefulness of the generated responses, reasons why they might have edited generated answers, and overall satisfaction using \aida.

\subsubsection{When were generated responses most useful?}

The instructor was asked about the relevance of the generated responses provided by \aida, to which they discussed the following scenarios when the responses were most useful.


\textbf{General questions:} For questions that the LLM likely already knew the answer to without additional context---typically technical matters relating to Java, compilation errors, and software development---\aida was extremely helpful in providing explanations and solutions.
Often the instructor would immediately use the \texttt{\#reply} prompt without searching for course-related context. Although they often attempted this without giving additional instructions (i.e., \texttt{\#reply$\varnothing$}), the instructor would usually resort to deleting the generated answer and re-attempt it by providing more specific guidance to the model (i.e., \texttt{\#reply$\blacksquare$}).

Despite the AI model being capable of understanding an underlying question, the instructor often had particular ideas on how they wanted  answers to be presented. They appreciated that the model could handle such generic questions without extra steps of narrowing down useful context, but still valued the ability to provide additional instructions on a per-question basis as they often changed from question to question. With that being said, the generated answers were generally perceived by the instructor as high-quality and did not require many modifications.

\textbf{Assignment specification questions:} Most of the questions on the discussion forum were related to the course assignments, and the instructor found that \aida was very helpful in generating answers to these questions. Despite the students being provided with detailed handouts and instructions, they often asked questions that were already answered in the handouts.
The instructor saw particular value in AIDA when students asked questions 
for which the instructor knew the answer was contained in their published resources, but could not immediately recall exactly where.
Even though the instructor developed the assignment resources, there were some situations where they could not recall specific details. 
Since \aida was able to efficiently locate relevant information from the handouts, the instructor found generating answers to these kinds of questions saved them a lot of time.


\textbf{Questions with little background provided:} Sometimes students asked questions without providing sufficient background or information that would allow the instructor to understand exactly what the issue might be. Instead of requesting the student to provide more details, the instructor found that using \aida to test the model's ability to generate a response was sometimes successful. Often the model could identify and propose solutions even when students supplied minimal and vague information. This was particularly useful to the instructor as it saved time and effort in requesting more information from the student.



\subsubsection{Reasons for large edits to generated answers}

The instructor was asked about the reasons as to why they edited some generated answers, especially when the content from some of the generated answers was nearly entirely deleted.


\textbf{Insufficient quality:} The instructor mentioned that the main reason for editing the generated answers was due to the quality of the responses. This is not limited to the correctness of the answer, but also the clarity and conciseness of the response. When re-attempting with further instructions still did not yield a satisfactory response, the instructor would often delete the bulk of the generated answer and write their own response. Fortunately, this was not a common occurrence.

\textbf{Simulating student responses:} When students interact anonymously on \ed, they are coded with a unique identifier that persists with their identity for that post (e.g., ``Anonymous Kangaroo'' or ``Anonymous Goose''). This allows students to refer to each other in a way that maintains their anonymity, but also allows them to distinguish between multiple fellow anonymous peers within the same post. Since \aida would often answer questions anonymously, their peers would reply to the bot as if it were a student and ask follow-up questions. There were times when the instructor did not need to utilise an LLM in the follow-up responses, but they needed to respond under the same anonymous identifier. In these cases, the instructor would action \texttt{[\#reply$\varnothing$ \#anon]} to generate the response entry, and then edit the content---often being a brief sentence to close the loop on the asking student's follow-up.

\subsubsection{Overall satisfaction and insights.}

Overall, the instructor found using \aida to be a positive experience; the generated answers were generally of high quality, and the instructor was satisfied with the responses provided by the model. The instructor found that using \aida could increase productivity, especially when the generated answers were of high quality and required minimal editing. Below they provide a few scenarios on \aida's convenience.


\textbf{Navigating context-driven responses:} The instructor found that when there were questions for which it was possible to find related contexts easily, high-quality answers could be generated by the LLM by incorporating these related contexts. However, when students asked general questions like ``\textit{I don't understand how interfaces work}''---which require code examples and explanations---it was very helpful and convenient to use \aida to generate those. The instructor also found that when they were asked a question that they themselves did not know the answer to, instead of manually performing an online search and reading through a range of posts, they would use \aida to give explanations to the question, as a replacement for doing an online search.



Several scenarios were identified where it was inconvenient for the instructor to respond using AIDA. When responses required a very specific context, determining which context to provide the model was cumbersome and sometimes led to trivial errors. The way that the most-likely contexts were displayed as lists when using the \texttt{\#help} prompt needed improvement to make it easier for the instructor to find the required information. Additionally, if the instructor intended to supply a brief answer, they sometimes found it quicker to write and submit directly rather than using AIDA.




\textbf{Ease of use of the prompts:}  In general, the instructor felt that the hashtags used in the prompts were intuitive and simple enough to memorise and use. The manner in which the hashtags and prompts could be entered into the regular comment box made it easy to use, and the instructor did not have any issues with the syntax of the prompts. This was especially the case as the \texttt{\#help} prompt would also provide a cheat sheet of the available prompts for the context results it returned.

With that said, the instructor noted that the \texttt{\#help} prompt needed improvement for easier information retrieval. It often provided an overload of information, including previous posts and related materials, making it challenging to identify the most relevant context to select. When a generated answer did not meet the instructor's expectations, they would return to the list of contexts and select alternative contexts, which was time-consuming. The instructor also noted that the \texttt{\#help} prompt worked well in test scenarios, but was somewhat less effective in diverse real-course situations. Previous posts sometimes only included questions and descriptions, lacking comprehensive answers. This further complicated context selection, especially if previous questions were unanswered or not correctly addressed.

\section{Contributions and Discussion}
In this practitioner paper, we described our experiences using AIDA, an instructor-driven tool that supports the generation of draft responses to student questions on online discussion forums.   AIDA worked well overall; most instructor edits to generated answers were minor, with deletions being the primary adjustment, indicating that the majority of AIDA-generated responses contained the necessary content. This result highlights the potential of instructor-in-the-loop AI systems for providing valuable, contextually accurate support to students while maintaining instructional oversight.

A key advantage of AIDA is its instructor-in-the-loop design, which is supported by  recent findings in educational AI tools such as Tutor CoPilot~\cite{wang2024tutorcopilothumanaiapproach}. Recent work using Tutor CoPilot demonstrated how Human-AI collaborative systems can support less experienced tutors by providing expert-like guidance, such as prompting student thinking rather than simply providing answers.
In a controlled trial with over 900 tutors, students whose tutors used Tutor CoPilot were more likely to master topics, with improvements especially pronounced for students taught by lower-rated tutors.
This underscores the effectiveness of tools that integrate teacher oversight rather than relying solely on automated guardrails, which can fail to adapt to real-time needs in educational settings.
A system like AIDA provides instructors the flexibility to approve or refine responses, to ensure alignment with course content and educational objectives.

Instructor feedback suggested that AIDA performed well in addressing general technical questions and specific assignment-related queries, and therefore achieved one of our goals which was to reduce the instructional burden on teaching staff which is a known concern related to moderating discussion forums.

\subsection*{Future Work}
To further improve AIDA's capabilities and the experience for instructors, we suggest several avenues that could be explored for future development:
\begin{itemize}

\item Expanding AIDA to handle multimodal inputs, such as images and screenshots, could address limitations in handling questions about visual debugging errors. Allowing AIDA to process and interpret visual information alongside text would help to improve response accuracy in cases that currently require manual intervention.

\item In our current implementation, AIDA relies on manual input from instructors to select and provide relevant contextual information for each response. This could be automated, so that AIDA dynamically selects and applies appropriate context as needed, improving efficiency and response accuracy without requiring as much instructor input.

\item Information about the student asking the question was not provided to AIDA in our current prototype, however tailoring responses based on the learning needs and background of the student posting the question could create a more personalised and effective experience. Adaptive responses could consider students' interaction histories or other information about their relevant knowledge.

\item  We would like to implement more robust metrics for assessing the impact of AIDA on student engagement, including a comparison of engagement levels before and after AIDA's introduction.  Metrics could include interaction frequencies and qualitative measures such as student feedback.

\item We provided AIDA with access to a rich repository of course content, including previous questions and answers.  However, these were all specific to our particular course. Providing the option to upload different resources would make it possible for AIDA to be easily customised for other courses.
\end{itemize}

\section{Conclusions}
In this work, we designed an AI bot that was integrated into the online discussion forum for a software engineering course.  Our goal was to support the instructor by generating draft responses for their review, which could be guided by relevant contextual information including course resources and posts and responses from previous iterations of the course.
We believe instructor-in-the-loop systems like AIDA, which allow teachers to have oversight into the AI-generated content that students see, represent a promising direction for the application of AI in education.  This design has shown that AI can effectively alleviate instructor workload by providing responses that often require minimal adjustments. The positive reception from both students and instructors underscores the value of human-centric AI systems in education.   Future work should explore multimodal capabilities, automated context selection, and personalised adaptive responses.






\bibliographystyle{ACM-Reference-Format}
\bibliography{references}


\end{document}